\documentstyle[aps,multicol,prb,epsfig]{revtex}

\input epsf

\begin{document}
\bibliographystyle{myprb1}
\title{Nonlinear $\sigma$ model for
 disordered superconductors}
\author{I. V. Yurkevich, and Igor V. Lerner}
\address{School of Physics and Astronomy, University of Birmingham,
Edgbaston, Birmingham B15 2TT, UK}

\date{24 January 2001}
\maketitle

\begin{abstract}
We suggest new variant of the nonlinear $\sigma$-model for the description
of disordered superconductors. The main distinction from existing models
lies in the fact that the saddle point equation is solved
non-perturbatively in the superconducting pairing field. It allows one to
use the model both in the vicinity of the metal-superconductor transition
and  well below its critical temperature with full account for the
self-consistency conditions. We show that the model reproduces a set of
known results in different limiting cases, and apply it for a
self-consistent description of the proximity effect at the
superconductor-metal interface.
\end{abstract}


\begin{multicols}{2}
\section{Introduction}

Since a seminal paper by Wegner, \cite{Weg:79}
a field theoretic approach to disordered systems based on the nonlinear
$\sigma$ model (NL$\sigma $M)
became one of the most powerful tools in describing
localisation effects and mesoscopic fluctuations.
The main advantage of this approach
lies in formulating the theory
in terms of low lying excitations (diffusion modes) which
greatly simplifies perturbative and renormalization group calculations, and
on the other hand allows a non-perturbative treatment.

Such an approach has been successfully extended to the description
of disordered superconductors.\cite{Fin:87,Fin:94,BeKi}
It was based on the Fermionic representation \cite{EfLKh}
of Wegner's NL$\sigma $M extended to include
the electron-electron interaction\cite{Fin}.
The starting point in these works \cite{Fin:87,Fin:94,BeKi} was a
microscopic model of interacting electrons in a random potential. The
effective NL$\sigma $M includes an extra Bosonic field describing
the superconducting order parameter $\Delta$.  Then
the lowest-order expansion in $\Delta$ is used. This makes
such an approach a good working tool in the vicinity of the superconducting
transition where all the interaction
channels can be easily included which makes  it very useful
in describing different aspects of the metal-superconducting transitions.

An alternative
 approach to the  NL$\sigma $M for dirty superconductors \cite{Op:87,%
KrOp,Kr:91,ASTs} starts from the Bogoliubov-de Gennes equations (or,
equivalently, Gorkov's equations)
without imposing a self-consistency condition on
the superconducting order parameter $\Delta$ which is considered as given.
Then the initial many-body problem turns into a single particle one
which makes applicable powerful techniques based on the
supersymmetric NL$\sigma $M. \cite{Ef:83}
Such a supersymmetric
 approach has been recently developed in Refs.\onlinecite{ASTs}
and applied to the description of non-perturbative aspects of the
proximity effect in superconducting--normal-metal structures. In this approach
 $\Delta$ was taken into account
just by the boundary conditions (Andreev reflection) for the normal region.
A natural disadvantage of this (and any supersymmetric) approach is that
no interaction can be included beyond the mean-field approximation;
thus it is impossible
to describe an  effect on the superconducting order parameter of
disorder in the normal metal (or even inside the superconducting region).

A novel NL$\sigma $M developed in this paper starts from a microscopic
model of electrons in a random potential with the BCS attraction, and
the  order parameter $\Delta$ is treated as a dynamical field, similar
to the earlier developed microscopic approach.\cite{Fin:87,Fin:94,BeKi}
We are using the standard fermionic replica approach \cite{EfLKh} in
temperature techniques.\cite{Fin} For a long time, it was widely believed
that such an approach cannot be used for non-perturbative analysis.
However, it was recently shown \cite{KamMez:99b,YL:99a} that this is not
the case, since the well-known exact non-perturbative result was reproduced
from the  fermionic replica   NL$\sigma$M, as well as more recently
\cite{AlKam}
within the  Keldysh technique.

In the initial  approach \cite{Fin}  to interactions within the  NL$\sigma$M,
a saddle point
approximation was identical to that of the non-interacting problem.
This scheme was recently greatly improved \cite{AnKam:99} by choosing
(within the  Keldysh technique) the saddle
point with taking account of the interaction which considerably simplified
any further analysis.  Such an analysis has been directly extended to dirty
superconductors
in Ref.~\onlinecite{FLS:00}.
We consider a model where, for simplicity, the Coulomb repulsion is not
included.  A distinctive feature  of our approach is
a change of the saddle point (and of a
subsequent initial approximation) in the presence of the superconducting
order parameter. This is similar but not identical to the choice suggested
in Ref.~\onlinecite{AnKam:99} (when applied to the Coulomb interaction, it would
lead to a different variant of the NL$\sigma$M).
The NL$\sigma$M \cite{AnKam:99} is optimized to maximally simplify the
lowest perturbational order while by sacrifizing this we arrive at quite
a general formulation of the model with different specific
approximations being made for different applications.

As usual, we restrict our consideration to the limit of dirty
superconductors when $\Delta\ll1/ \tau_{\text{el}}\ll \varepsilon_F$ (or,
equivalently, $v_F \tau_{\text{el}}\ll \xi$ where $\tau_{\text{el}}$
is the elastic mean free time, and $\xi$ in the correlation length in
dirty superconductors).
After describing describing in detail an alternative saddle-point approximation,
 we show how the model reproduces a set of known results
in different limiting cases, and apply it for a self-consistent
description of the proximity effect at the superconductor-metal interface.

\section{Basic Model}
We consider the standard BCS Hamiltonian in the presence of a random
potential, $u({\bf r})$. For completeness,
we start with outlining the standard procedure \cite{Fin:87}
of a field theoretic representation in the temperature technique
for this Hamiltonian. The corresponding action has the form
\begin{mathletters}
\begin{eqnarray}
S&=&S_0+S_i\\
S_0&=&\int\!{\rm d}x\,\psi_{s}^{*}(x)
\left[\frac{\partial}{\partial\tau}+
\label{xi}
\hat{\xi}+u({\mathbf r})\right]\psi_{s}(x) \\
\label{BCS}
S_i &=&\lambda_0\int\!{\rm d}x\,
\psi_{\uparrow}^{*}(x)\psi_{\downarrow}^{*}(x)
\psi_{\downarrow}(x)\psi_{\uparrow}(x),
\end{eqnarray}\end{mathletters}
Here $\psi_s(x)$ is a Grassmannian field\cite{Berezin,EfLKh}
anti-periodic in imaginary time $\tau$ with period $1/T$,
$x\equiv({\mathbf r},\tau)$, $s=(\uparrow,\,\downarrow)$ is
the spin index, $\lambda_0$ is the BCS coupling constant, and from now on
we set $\hbar=1$.

The random potential $u({\mathbf r})$ is  supposed to be  Gaussian with zero
mean and the standard pair correlator,
\begin{equation}
\left\langle u({\mathbf r})u({\mathbf r}\,')\right\rangle
=\frac{1}{2\pi\nu \tau_{\text{el}}}\,\delta ({\mathbf r}- {\mathbf r}\,')\,,
\end{equation}
with $\nu$ being density of states, and $\tau_{\text{el}}$
elastic mean free time.
The operator ${\hat \xi}$ in (\ref{xi}) is defined as
$$
{\hat{\xi}}
=\frac{1}{2m}\left(-i\nabla - \frac{e}{c}{\mathbf{A}}\right)^2-\mu\,,
$$
where $\bf A$ is a vector potential of an external magnetic field.

Averaging over  $u$ with the help of  the standard replica trick
gives the quartic in $\psi$ term in the action. Using the
Hubbard-Stratonovich transformation, one decouples both this
term and the BCS term, Eq.~(\ref{BCS}), the former
with the help of a matrix field $\hat\sigma =
\hat\sigma({\mathbf r};\tau,\tau')$ and the latter
with the help of a pairing field
$\Delta = \Delta({\mathbf r};\tau)$ which will eventually play the role
of the order parameter. This results in the following
effective action:
\begin{eqnarray}
{\cal S}[\hat\sigma,\Delta,\Psi]=
\frac{\pi\nu}{8 \tau_{\text{el}}}\,{\mathrm Tr}\, \hat\sigma ^2
+ \frac1\lambda_0\int\!{\rm d}x\, |\Delta(x)|^2+
\nonumber
\\[-1mm]
\label{PDS}
\\[-1mm]
\nonumber
+\int\!{\rm d}x\,\overline{\Psi}(x)\left[
-\hat\tau^{\text{tr}}_3\frac{\partial}{\partial\tau}-{\hat{\xi}}
+\frac{i}{2 \tau_{\text{el}}}\,\hat\sigma+ i{\hat \Delta}\right]\Psi(x)\,.
\end{eqnarray}
Here the replicated Grassmannian fields are
$$
\overline{\Psi} \equiv \left({\cal C}\Psi\right)^T= \frac{1}{\sqrt{2}}
\left(\psi^*_{s i}, -\psi_{s i}\right), \quad
\Psi^T = \frac{1}{\sqrt{2}}
\left(\psi_{s i},\psi^*_{s i}\right)\,,
$$
where $i=1\ldots N$ are the replica indices ($N=0$ in the final results). The
standard doubling of these fields ($\psi\to\Psi$) is convenient to separate
diffuson and cooperon channels for electrons propagating in the random
potential; $\cal C$ is the charge conjugating
matrix defined by the above equation.
The matrix fields ${\hat\sigma}$ and $\hat\Delta$ are defined in the space
spanned by $\Psi\otimes\overline\Psi$ which is convenient to think of as a
direct product of the $N\times N$ replica sector, $2\times 2$ spin sector, and
$2\times 2$ `time-reversal' sector.
The field $\hat\sigma$ is defined by its symmetries,
\begin{equation}
\hat\sigma^\dagger =\hat\sigma, \quad \hat\sigma= {\cal C}
\hat\sigma ^T {\cal C}^{-1}\,,
\label{sym}
\end{equation}
and Tr in Eq.\ (\ref{PDS}) refers to a summation over
all the matrix indices,  an integration over ${\mathbf
r}$ and a double integration over
 $\tau$ (as $\hat\sigma$ is not diagonal in $\tau$).

The field ${\hat \Delta}$ is an Hermitian and self-charge-conjugate
matrix field, which is diagonal in the
replica indices and coordinates $\mathbf r$ and $\tau$,
and has the following structure in the spin and time-reversal space:
\begin{equation}
{\hat \Delta} = -\left(\Delta '\,\hat\tau^{\text{tr}}_2 +\Delta '' \,
\hat\tau^{\text{tr}}_1
\right)
\otimes  \hat\tau^{\text{sp}}_2\,,
\label{Delta}
\end{equation}
where $\Delta '$ and $\Delta ''$ are real and imaginary parts of
the (scalar) pairing field
$\Delta$; $\hat\tau^{\text{tr}}_i$ and $\hat\tau^{\text{sp}}_i$ are Pauli
matrices ($i=0,1,2,3$ with $\hat\tau^{}_0=1$) that span the time-reversal and
spin sectors, respectively.

The integral over electron degrees of freedom is performed in a usual way,
so that one reduces the effective action (in
the  Matsubara-frequencies representation) to the following form:
\begin{eqnarray}
\nonumber
{\cal S}=\frac{\pi\nu}{8 \tau_{\text{el}}}\,{\mathrm Tr}\, \sigma ^2+
\frac{1}{T\lambda_0}\sum_{\omega}\int\!{\mathrm d}{\bf r}\,
|\Delta_{\omega}({\bf r})|^2
 \\
-\frac{1}{2}{\mathrm Tr}\ln\left[-{\hat{\xi}}
+ \frac{i}{2 \tau_{\text{el}}}\,\sigma
+ i\left(\hat{\epsilon}+ {\hat \Delta}\right)\right].
\label{a}
\end{eqnarray}
Here $\hat\epsilon={\mathrm diag}\,\epsilon_n$, while
 $\epsilon_n=\pi (2n+1)T$ is the
Fermionic frequency, and  $\omega=\epsilon-\epsilon'$ is the Bosonic
one. Since $\Delta$ is diagonal in the imaginary time $\tau$, it
is a matrix field in the Matsubara frequencies.

The action (\ref{a}) is a standard starting point for a further
field-theoretic analysis. To construct a working model, one needs to expand
in some way the Tr ln term in Eq.~(\ref{a}).
Our goal here is to derive a field theoretic model which is
fully self-consistent in terms of the superconducting order parameter
$\Delta$ and does not use a small-$\Delta$ expansion.
 We restrict our considerations to the limit of dirty
superconductors when $\Delta\ll1/ \tau_{\text{el}}\ll \varepsilon_F$.
Otherwise, we
do not impose any limitations on $\Delta$, and will derive the model
applicable both in the vicinity of the transition and deeply in the
superconducting regime.

\section{Saddle Point}

Our starting point is to construct a
saddle point approximation to the action
(\ref{a}) in the presence of the field $\hat \Delta$. As usual, we
 vary the action with respect to the field $\sigma$ which gives
\begin{equation}
\label{sp}
\sigma({\mathbf r}) = \left\langle{\mathbf r}\left|
\left[{-{\hat{\xi}}+ \frac{i}{2 \tau_{\text{el}}}\,\sigma+ i\left({
 \hat{\epsilon}} + {\hat \Delta}\right)}
\right]^{-1}\right|{\mathbf r}\right\rangle
\end{equation}
 As $1/\tau_{\text{el}}$ is much greater than both temperature $T$
and the order parameter $\Delta$, the matrix
$\hat\epsilon + {\hat \Delta}$ plays the role of a symmetry breaking field.
We look for a solution in a way similar to that in the metallic phase where
such a role is played by the matrix $\hat\epsilon $ alone.
In the metallic phase, the saddle-point equation with
$\epsilon\ne0$  has a unique solution $\hat\sigma=\Lambda$,
where $\Lambda $ is diagonal in
$\epsilon$ and unit in the replica and spin sectors:
\begin{equation}
\label{lambda}
\Lambda ={\text {diag\,\{sgn\,$\epsilon$\}}}\,,
\end{equation}
For $\epsilon=0$  a degenerate
solution to the saddle point equation is given by any
matrix $\hat\sigma $ of the symmetry (\ref{sym}) obeying the condition
$\sigma^2=1$. Such a matrix
can be represented as $\hat\sigma= U^\dagger  \Lambda
U$, with $U$ belonging to an appropriate symmetry group.\cite{GLKh}

Similarly, a solution to Eq.~(\ref{sp})
in the presence of $\hat\epsilon + {\hat \Delta}$ is given by
\begin{equation}
\hat\sigma_{s.p.} =
V_\Delta^\dagger  \Lambda V^{\,}_\Delta\,,
\label{s.p.}
\end{equation}
where  $V_\Delta$ is the matrix that simultaneously
 diagonalises both $\hat\sigma$ and
${{\hat\epsilon}} + {\hat \Delta}$. This means that it should be
found together with yet unknown eigenvalues $\lambda={\rm diag}\,
\lambda_\epsilon$ from
\begin{equation}
{\hat{\epsilon}} + {\hat \Delta} = V_\Delta^\dagger
\lambda^{\,}V^{\,}_\Delta\,.
\label{V}
\end{equation}
Naturally, one expects $V_\Delta$ to become unit matrix
above the superconducting transition temperature $T_c$.

Assuming that both the fields $\Delta(\bbox{r})$ and $\sigma(\bbox{r})$
are smooth functions of $\bbox{r}$ and looking for a spatially
independent solution to Eq.~(\ref{sp}) (i.e.\ ignoring at this
stage the fact that $\hat \xi$ and $V_\Delta$ do not commute),
one substitutes expressions (\ref{s.p.}) and
(\ref{V}) into Eq.~(\ref{sp}) thus reducing it to
\begin{equation}
\label{sp1}
\sigma = \left\langle{\mathbf r}\left|
\left[-{\hat{\xi}}+ \frac{i}{2 \tau_{\text{el}}}\,\Lambda+
i\lambda
\right]^{-1}\right|{\mathbf r}\right\rangle
\end{equation}
The scale of $\lambda$ is defined by $\epsilon\sim T$ and $\Delta$
which are both $\ll 1/\tau_{\text{el}}$ in a dirty
superconductor. Thus it is easy to verify that
the saddle point is given by Eq.~(\ref{s.p.})
with the eigenvalues $\Lambda$,  Eq.~(\ref{lambda}),
being  not affected by the presence of superconductivity.  Let us stress
that this saddle point is obtained by a  non-perturbative in
$\Delta$ rotation (\ref{s.p.}) of the metallic saddle point $\Lambda$.
This should lead to an
effective functional valid anywhere in the superconducting phase
rather than only in the vicinity of $T_c$.

Such an effective functional which includes
fluctuations around the saddle point is obtained in
the standard way.  First, one constructs
a saddle-point  manifold of matrices $\sigma$ obeying
the saddle-point equation at $\lambda=0$,
and then one expands the  Tr$\,$ln term in Eq.~(\ref{a})
in both the symmetry breaking term $\lambda$ and gradients of the
fields $V$.
The saddle-point manifold is convenient to represent as follows
\begin{equation}
\label{s}
\sigma= V_\Delta^\dagger
Q^{\,}V^{\,}_\Delta\,, \qquad Q= U^\dagger  \Lambda U\,,
\end{equation}
where $Q$ represents the saddle-point manifold in the
metallic phase and $\sigma$ is obtained from $Q$ by the same
rotation (\ref{s.p.}) as $\sigma_{s.p.}$ is obtained from the
metallic saddle point $\Lambda$.
Therefore, $Q$ is defined, as in the metallic phase, on the
coset space $S(2N)/S(N)\otimes S(N)$ where, depending on the
symmetry, $S$ represents the unitary, orthogonal or symplectic
group.  Before describing the expansion, let us stress that one
could expand the Tr$\,$ln term without making the rotation (\ref{s}),
i.e.\ in powers of $\nabla\sigma$ and of $\epsilon +\Delta$.
Although this would be formally an expansion within the same manifold,
performing first the rotation (\ref{s}) simplifies enormously all
the subsequent considerations and leads to a new variant of the
nonlinear $\sigma$ model.

 After substituting
 Eq.~(\ref{s}) into Eq.~(\ref{a}), one obtains the following
representation for the Tr$\,$ln term:
$$
\delta {\cal S} = -\case12 {\mathrm Tr}\ln \bigl\{\!\hat G_0^{-1}
+V^{\,}_
\Delta [\hat\xi,V^\dagger_\Delta]-i(U\lambda U^\dagger)
\bigr\}\,,
$$
where
$$
\hat G_0\equiv \left(
\hat \xi -\frac{ i}{2\tau_{\text{el}}}\Lambda
\right)^{-1}\,.
$$
The expansion to the lowest
powers of gradients and $\lambda$ is easily performed and results
after some straightforward calculations in the following
action:
\begin{eqnarray}
{\cal S}=
\frac{1}{T\lambda_0}\sum_{\omega}\int\!\!{\mathrm d}{\bf r}\,
|\Delta_{\omega}|^2 
+\frac{\pi\nu }{2}
{\mathrm Tr}\!\left[\frac D4\left(\partial
Q\right)^2
-
\lambda Q\right]\,,
\label{new}
\end{eqnarray}
where ${\mathrm Tr} $ refers to a summation over all the matrix indices and
Matsubara frequencies, as well as to an integration over $\bbox
r$. The long derivative in Eq.~(\ref{new}) is defined as
\begin{equation}
\partial Q
\equiv \nabla Q + \Bigl[{\bf A}_\Delta
\!\!-\!\frac{ie}{c} {\bf A}\hat\tau^{\text{tr}}_3, \,Q \Bigr]
\equiv \partial _0 Q + \left[{\bf A}_\Delta
, Q \right]
\,,
\label{LD}
\end{equation}
where the matrix ${\bf A}_\Delta$ is given by
\begin{equation}
\label{A}
{\bf A}_\Delta = V^{\,}_\Delta\partial_0 V^\dagger_\Delta  \,,
\end{equation}
and $\partial_0 $ is the long derivative (\ref{LD}) in the absence of the
pairing field $\Delta$.
Both $V^{\,}_\Delta$ and $\lambda$ should be found from the diagonalization
of $\epsilon + \Delta$, Eq.~(\ref{V}). Although such a diagonalization
cannot be done in general, it will be straightforward in many important
limiting cases.
For $\Delta=0$, the field ${\bf A}_\Delta$ vanishes, $\partial\to\partial_0$,
and $\lambda\to\epsilon$, so that the functional (\ref{new}) goes over to that
of the standard nonlinear $\sigma$ model for non-interacting
electrons.

The $\sigma$ model defined by Eqs.~(\ref{new})--(\ref{A})
is fully self-consistent, and the value of the
superconducting order parameter can be found from it for any
temperature and geometry (i.e.\ with a proper account of the
proximity effects, where applicable). The self-consistency condition would
easily follow from the variation of the action (\ref{new}) with respect to
$\Delta$ and finding the optimal configuration for the fields. However, it
is convenient to impose the
 self-consistency requirement only at the very end of calculations.
Any physical observable is then to be found by calculating an
appropriate functional average with the functional
(\ref{new})--(\ref{A}).

We proceed with illustrating how the model reproduces basic
fundamental results for dirty superconductors, then demonstrate how
to include consistently weak localization corrections in the
vicinity of the superconducting transition in the presence of a
magnetic field, and finally show how to take into account the self-consistency
of the order parameter
in the description of the proximity effect in the SNS geometry.

\section{The simplest approximation}

We show that the basic results for dirty superconductors can be
reproduced in the simplest approximation:
(i)
we neglect all nonzero Matsubara harmonics of the pairing field, i.e.\
substitute ${\hat \Delta}_0  \delta_
{\epsilon,-\epsilon'}$ for ${\hat \Delta}_{\epsilon\epsilon'}$;
(ii) we neglect
disorder-induced fluctuations near the saddle point, i.e.\
substitute the  saddle-point value $Q=\Lambda_\epsilon$.
In this case,
the matrix $\hat \epsilon + \Delta$
reduces to  direct product over all integer $n$ of
$(\hat\epsilon_n
+\hat\Delta_0)\otimes(\hat\epsilon_n
-\hat\Delta_0)$
where
\begin{equation}
\label{Delta0}
\hat\epsilon_n
+\hat\Delta_0\equiv
\left(
\begin{matrix}{
\epsilon_n&\Delta_0\cr
 \Delta^*_0 &-\epsilon_n
}
\end{matrix}
\right)\;.
\end{equation}
Here $\Delta_0=|\Delta|\,{\mathrm e}^{i\chi}$ is a
two-component field which, naturally, plays the role of the order
parameter (we omit the index 0 in $|\Delta|$).
Now it is easy to find explicitly the eigenvalues $\lambda$
and the diagonalising matrix $V_\Delta$ in Eq.~(\ref{V}).
\begin{eqnarray}
\lambda_\epsilon &=&
\sqrt{\epsilon_n^2   + |\Delta|^2}\,{\mathrm sgn}\, \epsilon_n \,,\quad
\cos\theta_\epsilon\equiv\epsilon_n/\lambda_\epsilon
\nonumber
\\[-1mm]
\label{diagonal}
\\[-1mm]
\nonumber
V_{n\Delta}({\bf r})
&=& \cos\frac{\theta_\epsilon}2+\hat\delta\,{\rm sgn}\,\epsilon_n\,
\sin\frac{\theta_\epsilon}{2}\,.
\end{eqnarray}
where $\hat\delta\equiv(\hat \Delta_0/|\Delta|)\delta_{\epsilon, -
\epsilon'}$ is the $4\times4$ matrix which depends only on the
phase $\chi$ of the field $\Delta_0$ and repeats the matrix
structure of $\hat \Delta_0$, Eq.~(\ref{Delta}),
and the full matrix $V_{\Delta}$ is the direct product of all $V_{n\Delta}$.

On utilizing the assumption (ii) above, i.e.\ $Q=\Lambda$,
and substituting the parameterisation (\ref{diagonal})
into the Eq. (\ref{new}),  we arrive at the action ${\cal S}
\equiv \int\!\!{\mathrm d}^dr\, {\cal L}$ with
\begin{eqnarray}
{\cal L}
&=&\frac{|\Delta|^2}{\lambda_0 T}
- 2\pi\nu\sum_{{\epsilon}}\sqrt{{\epsilon}^2 + |\Delta|^2}
+\delta {\cal L}\,,
\nonumber
\\[-3mm]
\label{L}
\\[-1mm]
\nonumber
\delta{\cal L}&\equiv&\frac{\pi\nu D}{2}\sum_{{\epsilon}}\left[
\left(\nabla\theta_{{\epsilon}}\right)^2 +
\sin^2\theta_{{\epsilon}}\left(\!\nabla\chi-\frac{2e}{c}{\bf A}\!\right)^{\!2}
\right]\,.
\end{eqnarray}
Using the parameterization (\ref{diagonal}) one can easily sum
over ${\epsilon}$ to get
\begin{equation}
\label{19}
\delta{\cal L}=\frac{\pi\nu D}{8T}\left\{C_1 (\nabla |\Delta|)^2
+ C_2\!\left(\!\nabla\chi -\frac{2e}{c}{\bf A}\!\right)^{\!2}
\right\}\label{F}
\end{equation}
where the stiffness coefficients $C_{1,2}$ are given by
\begin{eqnarray}
C_1 &=& \frac{1}{|\Delta|}\tanh\frac{|\Delta|}{2T}
+\frac{1}{2T}\cosh^{-2}\frac{|\Delta|}{2T}\,,
\nonumber
\\[-1mm]
\label{C}
\\[-1mm]
\nonumber
C_2 &=& 2|\Delta|\tanh\frac{|\Delta|}{2T}\,.
\end{eqnarray}
The functional (\ref{F})--(\ref{C}) coincides with that obtained in
Ref.\onlinecite{Kr:91}.
Expanding coefficients $C_{1,2}$ in $\Delta$, one obtains the
  Ginzburg-Landau functional as that in Ref.\onlinecite{Kr:91}. However,
the simplest approximation used here (and equivalent to those on which
earlier considerations\cite{Op:87,KrOp,Kr:91} were based) is not sufficient
even in describing the vicinity of the superconducting transition. In general,
one must keep all the Matsubara components of the pairing fields. In the
following section, we will show how to do this in the  vicinity of the
transition in the weak disorder limit.

\section{Ginzburg-Landau Functional}

In the vicinity of the superconducting transition one can expand the action
(\ref{new}) in the pairing field. A further simplification is possible by in
the weak disorder limit, $p_F\ell\gg1$: one can integrate out the $Q$-field to
obtain an effective action for the $\Delta$-field only. In the quadratic in
$\Delta$ approximation, the kernel of this action will give an effective
matrix propagator of the pairing field, with due account for the disorder,
which governs properties of a disordered superconducting sample near the
transition.

To integrate over the $Q$-field, one splits the action (\ref{new}) into ${\cal
S} \equiv {\cal S} _0 + {\cal S} _\Delta$ where
\begin{equation}
\label{FinN}
{\cal S} _0=-\frac{\pi\nu D}{8}{\mathrm Tr}\,(\partial _0 Q)^2
-\frac{\pi\nu}{2}{\mathrm Tr}\, \epsilon Q ,
\end{equation}
is the standard nonlinear $\sigma$ model functional as in the metallic phase.
Then one makes a cumulant expansion, i.e.\ first expands ${\rm e}^{-({\cal S}
_ 0+ {\cal S} _\Delta)}$ in powers of ${\cal S} _\Delta$, then performs the
functional averaging with ${\rm e}^{- {\cal S} _0}$ (denoted below by
$\langle\ldots\rangle_Q$) and finally re-exponentiates the
results. The expansion involves only the first and second order cumulants
since the higher order cumulants generate terms of higher order in $\Delta$.
Then the only terms which contribute to the action quadratic in $\Delta$ are
given by
\begin{eqnarray}
\lefteqn{{\cal S}_{\rm eff}[\Delta] =\frac{1}{\lambda_0 T}\sum_{\omega}\int
\!\!{\mathrm d}{\bf r}\,
|\Delta_{\omega}|^2 -\frac{\pi\nu}{2}\Bigl\langle{\mathrm Tr}\, (\lambda
\!-\!\epsilon) Q\Bigr\rangle_{\!Q} }
\nonumber
\\[-1mm]
\label{av}
\\[-1mm]
\nonumber
&&-\biggl\langle\frac{\pi\nu D}{8}{\mathrm Tr}\,\left[{\bf
A}_\Delta,Q\right]^2 +\frac{(\pi\nu D)^2}{8}\Bigl( {\mathrm Tr}\, Q\partial _0
Q {\bf A}_\Delta \Bigr)^{\!2}\biggr\rangle_{\!\!Q}.
\end{eqnarray}
Expanding $\lambda$ and ${\bf A}_\Delta$ to the lowest power in $\Delta $ and
performing a standard functional averaging, as described in Appendix, one
finds the action quadratic in $\Delta$ as follows:
\begin{equation}
{\cal S}_{\rm eff}[\Delta] =\frac{\nu}{T}\sum_{\omega}\int\!{\rm d} {\bf r}\,
\Delta^*_{\omega}({\bf r})\bigl<{\bf r}\bigl|\,\hat{\cal K}_{\omega}
\,\bigr|{\bf r}'\bigr> \Delta_{\omega}({\bf r}')\,,
\label{K}
\end{equation}
with the operator $\hat{\cal K}_{\omega}$ given by
\begin{equation}
\label{wl}
\hat{\cal K}_{\omega}=\frac{1}{\lambda_0 \nu} -2\pi T\!\!\!\!
\sum_{\epsilon(\omega-\epsilon)<0} \left\{ \hat\Pi^{c}_{\omega}
+\frac{1}{\pi\nu} \frac{
\Pi^{d}_{|2\epsilon-\omega|}(0){\hat {\cal C}}}{(2\epsilon-\omega)^2} \right\}.
\end{equation}
Here $\Pi^{c,d}_{|\omega|}({\bf r}, {\bf r}') =\bigl<{\bf
r}\bigl|\hat\Pi^{c,d}\bigr|{\bf r}'\bigr> $ are the cooperon and diffuson
propagators, respectively, with
\begin{equation}
\label{Pi}
\hat\Pi^{c}_{|\omega|} = \Bigl( {\hat{\cal C}} + |\omega| \Bigr)^{-1}\,,
\end{equation}
where the operator $ {\hat {\cal C}}\equiv -D\left(\nabla - {2ie}{\bf
A}/{c}\right)^2$ defines the propagation of the cooperon modes;
$\hat\Pi^{d}$ is obtained from $\hat\Pi^{c}$ by putting $\bf A=0$.

In the last term in Eq.~(\ref{wl}),
$\Pi^{d}_{|\omega|}
(0)\equiv \Pi^{d}_{|\omega|}({\bf r},{\bf r})$;
this term may be obtained by expanding (in the weak
disorder parameter)
the cooperon propagator  with the renormalized diffusion
coefficient,
$$
{\hat{\cal C}} \to
\left[1-\frac{1}{\pi\nu}\Pi^{d}_{|\omega|}(0)\right]\,{\hat{\cal C}}.
$$
Therefore, this is just
a weak localisation correction to the free cooperon propagator
$\Pi^{c}_{|\omega|}({\bf r}, {\bf r}')$.

The summation over Matsubara frequencies in Eq.~(\ref{wl}) is
easily performed to yield
\begin{equation}
\hat{\cal K}_{\omega}=
\ln\frac{T}{T_0}+\psi\!\!\left(\!
\frac{1}{2}+\frac{|\omega|\!-\!{\hat{\cal C}} }{4\pi
T}\!
\right)
-\psi\!\!\left(\frac{1}{2}\right) - \frac{a_{\omega}{\hat{\cal C}}}{4\pi T}
\,,
\label{K1}
\end{equation}
where $T_0\equiv T_{c0}(B\!=\!0)$ is the transition temperature of the clean
superconductor in the absence of a magnetic field. The weak localisation
correction is proportional to the coefficient $a_{\omega}$
given by
$$
\begin{array}{l}
\displaystyle
a_{\omega}(T)=\frac{1}{\pi\nu V}\sum_{\mathbf{q}}\frac{1}{Dq^2}\left\{
\psi'\left(\frac{1}{2}+\frac{|\omega|}{4\pi T}\right) \right. \\[6mm]
\displaystyle
\left. -\frac{4\pi T}{Dq^2}
\left[
\psi\left(\frac{1}{2}+\frac{|\omega|+Dq^2}{4\pi T}\right)
-\psi\left(\frac{1}{2}+\frac{|\omega|}{4\pi T}\right)
\right]
\right\}.
\end{array}
$$
For $\omega=0$ the coefficient $a_0\equiv a_{\omega =0}(T)$
can be simplified in the two limits:
\begin{equation}
 a_0=\left\{
\begin{array}{ll}
\displaystyle
\quad\frac{\psi'({1/2})}{\pi\nu L^d}\!\!\!\sum\limits_{L_{T}^{-1}
<q<\ell^{-1}}\frac{1}{Dq^2}\,, &
L \gg L_T\,,\\[8mm]
\displaystyle
-\frac{\psi''({1/2})}{8\pi^2\nu L^d T}\,\,, &
L \ll L_T \,,
\end{array}\right.
\end{equation}
where $L_T\equiv\sqrt{D/T}$ is the thermal smearing length.

The instability of the normal state (i.e. a transition into the
superconducting state)
occurs when the lowest eigenvalue of the operator ${\hat{\cal K}}_{\omega}$
becomes negative. The eigenfunctions of this operator coincide with
the eigenfunctions of the cooperon operator ${\hat{\cal C}}$. The
lowest eigenvalue of ${\hat{\cal C}}$
 is known to be ${\cal C}_0= DB/\phi_0$, where
$\phi_0$ is the flux quanta. This ground state cooperon eigenfunction
corresponds to the lowest eigenvalue ${\cal K}_0$ of the operator
${\hat{\cal K}}_{\omega}$. The condition ${\cal K}_0=0$ implicitly
defines the line $T_c(B)$ in the $(T,B)$-plane where the transition occurs:
\begin{equation}
\ln\frac{T_c}{T_0} +\psi\left(\frac{1}{2}+\frac{{\cal C}_0}{4\pi T_c}\right)
-\psi\left(\frac{1}{2}\right)=\frac{a_0 {\cal C}_0}{4\pi T_c}.
\label{Tc(B)}
\end{equation}
The term in the r.h.s.\ of Eq.~(\ref{Tc(B)})
 describes a $1/g$-correction to the main result. This
 weak localisation is linear in the magnetic field $B$ and
vanishes as $B\to 0$ as expected (Anderson theorem). In a nonzero magnetic
field the weak localisation
correction to the $B_c$ is positive which has a very simple explanation.
The superconductivity is destroyed by the magnetic field when the flux over
the area with the linear size of the order of the coherence length becomes
greater than the flux quanta. The weak localisation corrections diminish the
diffusion coefficient which leads to the shrinkage of the coherence length.
Therefore, one needs more strong field to fulfill the condition of the coherence
destruction. The same reasoning explains the growth of $T_c$ in the fixed
magnetic field.

Note finally that we have calculated the $Q$-averages in Eq.~(\ref{av})
perturbatively, up to the first order in the weak-localization correction.
It would be
 straightforward to include main weak-localization corrections in all orders
by calculating these averages via the renormalization group. This
would lead to the renormalizing the diffusion coefficient in the cooperon
propagator (\ref{Pi}), thus changing the shape of the $T_c(B)$ curve.
However, the value of $T_c(0)$ will again remain unaffected, since
the superconducting instability is defined by the appearance of
 the zero mode in the
operator $\hat {\cal K}$,  Eq.~(\ref{K1}).  This zero mode is homogeneous, and
thus does not depend on the value of the diffusion coefficient in the cooperon
propagator.

\section{Proximity Effect}
A recent supersymmetric version\cite{ASTs} of the NL$\sigma $M has been
specifically formulated for studying the proximity effect in SNS junctions.
Although this version is very convenient for a non-perturbative analysis,
it has a natural disadvantage of the supersymmetric approach: no interaction
can be included beyond the mean-field approximation. It means that the
superconducting order parameter $\Delta$
should be treated as a background field rather than a dynamical one.
More specifically, $\Delta$ was taken into account \cite{ASTs}
just by the boundary conditions (Andreev reflection) at the boundaries of a
normal metal, while having been considered as a given field in the
superconducting region. This allows for changes in characteristics of the
normal metal in the proximity of the superconductor, but not for the possibility
of changes in the superconducting order parameter in the proximity of the
normal metal.

The action in the normal region ($N$) has the standard form\cite{EfLKh,AKL:86}
while in the superconductor ($S$) we have NL$\sigma $M of the
form Eq.(\ref{new}).
The continuity of the Green function across the $N/S$
boundary requires

\begin{equation}
\label{bc}
\left. Q_N \right|_{N/S}
= \left. V_{\Delta}^\dagger \, Q_S V_{\Delta} \right|_{N/S}.
\end{equation}
The
$N$ region by itself would favour $Q_N = \Lambda$. The
proximity leads to
 rotation of matrix  $Q_N$ in the $N$ region
in order to match the structure imposed by the boundary condition (\ref{bc})
\begin{equation}
Q_N \longrightarrow V_N^\dagger \, Q_N  \, V_N \,,
\end{equation}
with the rotation matrix $V_N$ of the same structure as
$V_{\Delta}$ in the $S$ region  so that
at the boundary they match each other.
Proceeding in the same manner as above we keep
only the $\omega=0$ component
 of the pairing field and neglect the disorder induced fluctuations,
i. e. we put $Q_N=Q_S=\Lambda$. Then for the $N$ region we have

\begin{equation}
\begin{array}{l}
\displaystyle
Q \to V_N^\dagger \,\Lambda \, V_N
=\cos\theta_{{\epsilon}}+\sin\theta_{{\epsilon}}\,
{\hat{\bf \delta}}, \\[6mm]
\displaystyle
{\hat{\bf \delta}}_{\epsilon,\epsilon'}
= -\delta_{\epsilon,-\epsilon'}\left(\cos\chi_{{\epsilon}}\,
\hat\tau^{\text{tr}}_2 +\sin\chi_{{\epsilon}}\hat\tau^{\text{tr}}_1\right)
\otimes i\sigma_2
\end{array}
\end{equation}
where $\theta_{{\epsilon}}$ and $\chi_{{\epsilon}}$ are now independent
variables. In a bulk superconductor, all these parameters were explicit
functions of $\Delta$ and $\epsilon$,  Eq.~(\ref{diagonal}).
There is no such a constraint in the normal region. The
($\epsilon,-{\epsilon}$) sectors in the normal region are still
coupled due to proximity effect
but they may all be different.

In this approximation the action corresponding to the $N$ region
decouples into the sum of uncorrelated contributions:
\begin{eqnarray}
\label{n}
{\cal S}_N &=& 2\pi\nu\sum_{{\epsilon}}\int\!{\mathrm d}{\bf r}\,
{\cal L}_\epsilon \,,\\
\nonumber
{\cal L}_\epsilon&=&
\frac{D}{4}\left[(\nabla\theta_{{\epsilon}})^2
+ \sin^2\theta_{{\epsilon}}\left(\nabla\chi_{{\epsilon}}
-\frac{2e}{c}{\bf A}\right)^2\right]
- \epsilon\cos\theta_{{\epsilon}}
\end{eqnarray}
Now
we find supercurrent ${\bf j}_s$
 by varying the action (\ref{n}) with respect to the
vector potential $\bf A$:

\begin{equation}
\label{supercurrent}
{\bf j}_s = 2e \pi\nu D\,\, T\, \sum_{\epsilon}
\Bigl\langle\sin^2\theta_{\epsilon}
\Bigl(\nabla\chi_{\epsilon}-i\frac{2e}{c}{\bf A}\Bigr)\Bigr\rangle_N,
\end{equation}
where $\langle\ldots\rangle_N$
 stand for the functional  averaging with the action Eq.(\ref{n}),
the functional integration being performed
 over the functions obeying the boundary conditions
\begin{equation}
\left.\chi_{{\epsilon}}\right|_{N} = \left.\chi\right|_{S}, \quad
\left.\cos\theta_{{\epsilon}}\right|_{N}
= \frac{{\epsilon}}{\sqrt{{\epsilon}^2+|\Delta|^2}}\,.
\end{equation}
Here $|\Delta|$ and $\chi$ are modulus and phase of the
order parameter at the $N/S$ interface.

The classical trajectory corresponding to the action (\ref{n}) is nothing but
the Usadel equation:\cite{Usadel}

\begin{equation}
\label{usadel}
\begin{array}{l}
\displaystyle
-\frac{D}{2}\Delta\theta
+\frac{1}{4} \sin^2\theta\left(\nabla\chi-\frac{2e}{c}{\bf A}\right)^2
+ \epsilon\cos\theta = 0, \\[4mm]
\displaystyle
\nabla\left[\sin^2\theta\left(\nabla\chi-\frac{2e}{c}{\bf A}\right)\right]=0.
\end{array}
\end{equation}

For quasi-1$D$ geometry in the absence of a magnetic field,
the Usadel equation (\ref{usadel}) can be written as
the equation for $\theta$

\begin{equation}
-\frac{{\mathrm d}^2\theta}{{\mathrm d} x^2}
+ \alpha_{\epsilon}^2\frac{\cos\theta}{\sin^3\theta}
+ L^{-2}_{\epsilon}\sin\theta = 0\,,
\end{equation}
with the self-consistency condition on $\alpha_{\epsilon}$

\begin{equation}
\label{chiN}
\chi_N = \alpha_{\epsilon}\int\frac{{\mathrm d} x}{\sin^2\theta_{\epsilon}}\,.
\end{equation}
Here $\chi_N\equiv \chi_+ - \chi_-$ is the phase difference
between two superconducting banks and $L_{\epsilon}=\sqrt{D/2\epsilon}$
is the coherence length for two particles with the energy difference $\epsilon$
propagating in the normal metal. For a long normal bridge between the
 two superconducting banks,
$L\gg L_T\equiv\sqrt{D/2\pi T}$,  one may consider separately
three regions: those close
to the
$N/S$ boundaries (with the width of order  $L_T$) and the bulk. Matching
the solutions for all the regions,
we find the following expression which well approximates the solution for
the entire normal region:

\begin{figure}
\epsfxsize=.9\hsize \epsffile{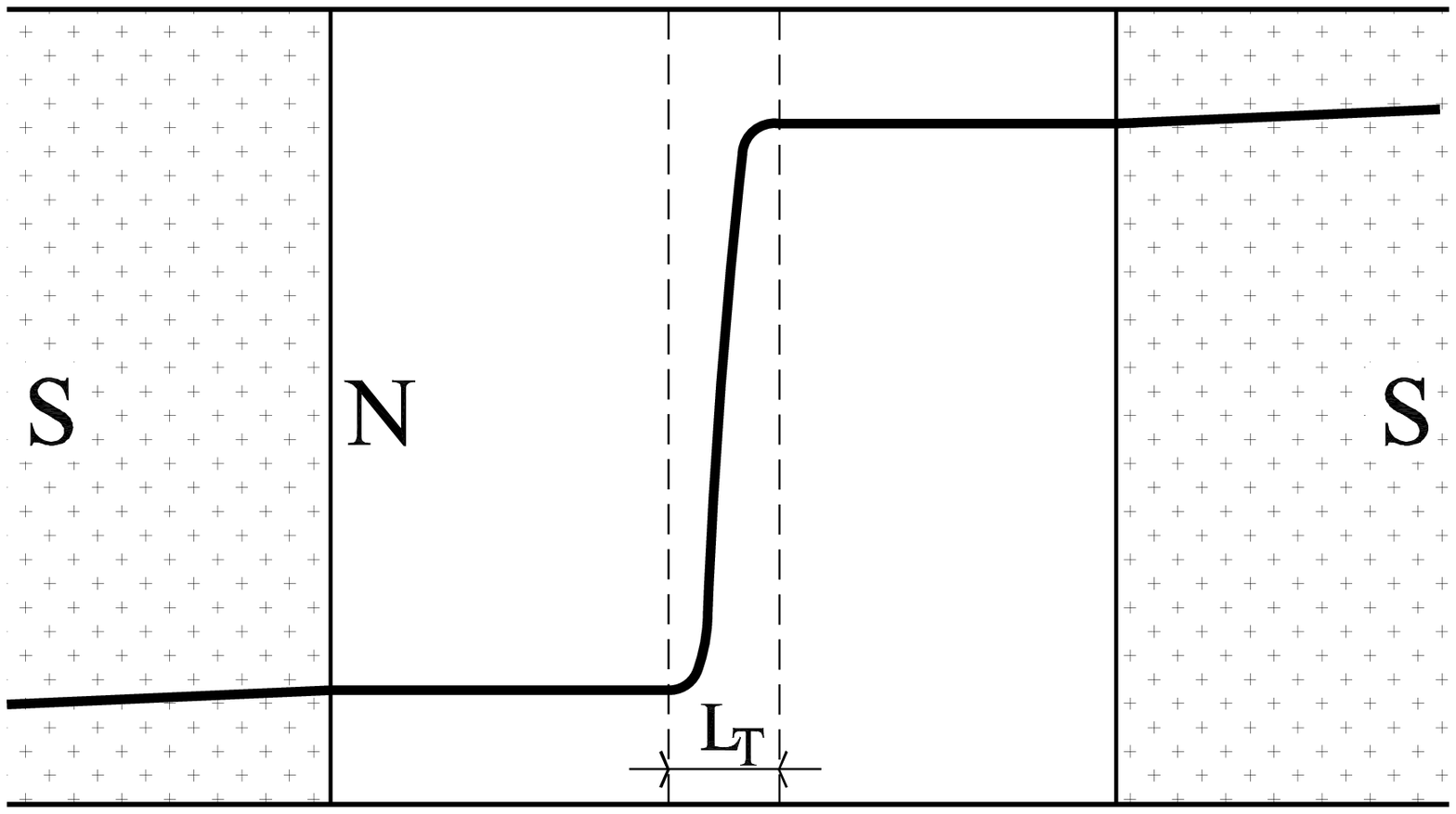}

{\small \setlength{\baselineskip}{10pt} FIG.\ 1.
 A spatial dependence of  the phase $\chi_\varepsilon $
 across the SNS contact for quasi-1D geometry. }
\end{figure}


\begin{equation}
\begin{array}{l}
\displaystyle
\theta(x) = 8\tan(\theta_0/4)\, e^{-L/2L_{\epsilon}}
\sqrt{\cos^2\frac{\chi_N}{2}+\sinh^2\frac{x}{L_{\epsilon}}}\,,\\[5mm]
\alpha_{\epsilon}=32\tan^2(\theta_0/4)\,\sin\chi_N\,L_{\epsilon}^{-1}
\exp{\left[-L/L_{\epsilon}\right]}\,,
\end{array}
\end{equation}
where $\left.\theta_0\equiv\theta_{\epsilon}\right|_{N/S}$. In calculating
the supercurrent through the normal bridge, one reduces the expression within
the angular brackets in (\ref{supercurrent}) to
$ \sin\chi_N  \alpha_{\epsilon}
$.
Then it is enough to keep only the leading term with $\epsilon_0=\pi T$ because
the contributions from  all the other frequencies are exponentially suppressed
as
$L_{\epsilon}<L_T$. Then
we obtain the following expression typical for the Josephson junctions
$j_s = j_c\,\sin\chi_N$,
where $j_c$ is the critical current:

\begin{equation}
j_c = e 2^7\pi\nu D\,\, T\,
\tan^2(\theta_T/4)\,L_{T}^{-1}
\exp{\left[-L/L_{T}\right]}\,,
\end{equation}
with $\theta_T\equiv\theta_{\epsilon_0}$.

The supercurrent in the  superconducting banks is found by varying the action
(\ref{19}) valid in the $S$ region with respect to the
vector potential $\bf A$:

\begin{equation}
\displaystyle
{\bf j}_s=e\pi\nu D |\Delta| \tanh\frac{|\Delta|}{2T}\,\frac{\chi_S}{L_S},
\end{equation}
where $L_S$ is the length of the superconductor and $\chi_S$ the phase
difference between its edges.

It should be stressed that we have varied the action
for the entire SNS structure, rather than only for the normal region
subject to the boundary conditions at the superconducting banks as in the
supersymmetric variant of the NL$\sigma$M for dirty superconductors.
\cite{ASTs}
This means that the  the phase difference across the normal region is not fixed
but should be found self-consistently
by finding the  optimal
configuration  for the action for the entire  SNS structure subject by the
matching the fields at the $N/S$ boundaries.
This defines the actual phase difference $\chi_N$, Eq.\ (\ref{chiN}),
 across the normal bridge. Numerically, a similar procedure has been employed
in Ref.~\onlinecite{RCB:96}.
It is easy to show that the matching
condition can be expressed as the continuity
of the supercurrents (as varying with respect to the phase difference is
equivalent to varying with respect to the vector potential).
Thus the supercurrent conservation defines
the phase difference on the normal bridge:
\begin{equation}
\frac{|\Delta|}{64 T} \tanh\frac{|\Delta|}{2T}\,\chi_S
=
\frac{L_S}{L_{T}}
{\rm e}^{-{L}/{L_{T}}}\,
\sin\chi_N \tan^2\frac{\theta_T}4,
\end{equation}
so that if the width of superconductor banks $L_s$ is sufficiently large,
the overall phase drop mainly  happens across the banks.

Finally, let us reiterate that the main result of the paper is a novel
variant of the NL$\sigma $M given
by Eqs.~(\ref{new})--(\ref{A}).
Here we have applied this formalism to a few relatively simple problems mainly
to show that it works and has certain advantages over alternative variants
of the NL$\sigma $M. This model has also been applied to a microscopic
consideration \cite{YL:00} of the quantum phase slip problem in quasi-1D
superconductors,\cite{LanAmb,Duan}
 and to a microscopic derivation of level statistics in
nonstandard symmetry classes introduced in Ref.~\onlinecite{AZ:97}.
Let us also stressed that the method employed in the derivation of
Eqs.~(\ref{new})--(\ref{A}) can be straightforwardly generalized both to
including different types of interactions, and to considering the
unconventional pairing in dirty superconductors.

\acknowledgments
We are grateful to R.~A.~Smith for useful comments.
 This work has been supported by the Leverhulme Trust under Contract No.\ 
F/94/BY.

\begin{appendix}
\section*{}
To perform the functional
averaging in Eq.~(\ref{av}), one should employ some parameterization of the
field $Q$ in terms of unconstrained matrices, for example\cite{Ef:83,AKL:86}
$$
Q=(1-W/2)\Lambda (1+W/2)^{-1}\,,
$$
where $W=-W^\dagger$, and $W\Lambda  + \Lambda  W=0$.
The $Q$-integration then reduces to the Gaussian
one with the weight ${\rm e}^{-S_0}$ with $S_0$ obtained from
Eq.~(\ref{FinN}) by expanding $Q$ to the second order in $W$.
The Gaussian $W$ integration is carried out with the help of
the following contraction rules
\begin{eqnarray}\nonumber
\lefteqn{\Bigl<
{\rm Tr} M\,W({\bf r})\, P\,W({\bf r}')\Bigr>
=-\frac{2}{\pi\nu}
\times}
\\[-1mm]
\label{lt}\\[-1mm]\nonumber
&&
\sum\limits_{
\begin{array}{c}
\epsilon\epsilon'<0\\
\alpha,\beta
\end{array}}
\!\!\!\!\Bigl[
\left({\hat \pi}
\,\tau_1\right)^{\alpha\beta}_{\epsilon\epsilon'}
{\mathrm tr} M^{\alpha\beta}_{\epsilon\epsilon'}\,
{\bar P}^{\beta\alpha}_{\epsilon'\epsilon}
+
{\hat \pi}^{\alpha\beta}_{\epsilon\epsilon'}
{\mathrm tr} M^{\alpha\alpha}_{\epsilon\epsilon}
{\mathrm tr} P^{\beta\beta}_{\epsilon'\epsilon'}
\Bigr]
\end{eqnarray}
\begin{eqnarray}\nonumber
\lefteqn{\!\!\!\!
\left\langle\,{\rm Tr} M W ({\bf r}) {\rm Tr} P W({\bf r}')\right\rangle
=- \frac{2}{\pi\nu}\times}
\\[-1mm]
\label{st}\\[-1mm]\nonumber
&&
\sum\limits_{
\begin{array}{c}
\epsilon\epsilon' <0\\
\alpha,\beta
\end{array}}
{\hat \pi}^{\alpha\beta}_{\epsilon\epsilon'}\,
{\mathrm tr}
(M-{\bar M})^{\alpha\beta}_{\epsilon\epsilon'}\,
(P-{\bar P})^{\beta\alpha}_{\epsilon'\epsilon}
\end{eqnarray}
where the upper indices $\alpha,\, \beta$ refer to the time-reversal
 sector and tr refers only to  the matrix
indices which are not indicated explicitly.
The matrix ${\hat \pi}$ in the
Eqs.(\ref{lt}) and (\ref{st}) has the following structure in the time-reversal
sector:

\begin{equation}
{\hat \pi}_{\epsilon\epsilon'}({\bf r},{\bf r}') = \left(
\begin{array}{ll}
\displaystyle
\Pi^d_{|\epsilon-\epsilon'
|}({\bf r},{\bf r}') & \Pi^c_{|\epsilon-\epsilon'
|}({\bf r},{\bf r}') \\
\Pi^c_{|\epsilon-\epsilon'
|}({\bf r}',{\bf r}) & \Pi^d_{|\epsilon-\epsilon'
|}({\bf r},{\bf r}')
\end{array}\right)
\end{equation}
 where the propagators are solutions to the standard cooperon and diffuson
equations:
\begin{equation}
\begin{array}{l}
\displaystyle
\left[-D\nabla_{{\bf r}}^{2} + \omega\right]\Pi_{\omega}^{d}({\bf r},{\bf r}')
=\delta ({\bf r}-{\bf r}'), \\[6mm]
\displaystyle
\left[-D(\nabla_{{\bf r}}-i\frac{2e}{c}{\bf A}({\bf r}))^2
+ \omega\right]\Pi_{\omega}^{c} ({\bf r},{\bf r}')
=\delta ({\bf r}-{\bf r}').
\end{array}
\end{equation}
Note that in the absence of magnetic field these contraction rules go over to
those previously derived for the orthogonal symmetry. \cite{IVL:92}

Next, one expands
$Q$ in Eq.~(\ref{av}) up to the 4th power in $W$ and uses
the above contraction rules to obtain

\begin{eqnarray}
\left\langle {\rm Tr} (\lambda - \epsilon ) Q \right\rangle_{\!\!Q}
&=& {\rm Tr} (\lambda - \epsilon ) \Lambda\,,
\\[5mm]
\displaystyle
\left\langle
{\mathrm Tr}\,\left[{\bf
A}_\Delta,Q\right]^2 \right\rangle_{\!\!Q} &=&
{\mathrm Tr}\,\left[{\bf
A}_\Delta,\Lambda\right]^2 \cr
+ \frac{8}{\pi\nu}\sum_{\epsilon(\omega-\epsilon)<0}& &
\frac{\Pi^d_{|2\epsilon-\omega|}(0)}{(2\epsilon-\omega)^2}
\left|(\nabla-\frac{2e}{c}{\bf A})\Delta_{\omega}\right|^2
\end{eqnarray}
Taking into account that the second term in the brackets in the
Eq.~(\ref{av}) contributes to the higher order correction only we
arrive at the Ginzburg-Landau functional described in the text.

\end{appendix}


\end{multicols}
\end{document}